\documentclass[preprint,aps,showpacs]{revtex4}
\usepackage{graphicx}
\begin{document}

\title{Sensitivity to multi-phonon excitations in heavy-ion fusion reactions}
\author{H. Esbensen}
\affiliation{Physics Division, Argonne National Laboratory, Argonne, Illinois 60439}
\date{\today}

\begin{abstract}
Measured cross sections for the fusion of $^{64}$Ni with $^{64}$Ni, $^{74}$Ge, 
and $^{100}$Mo targets are analyzed in a coupled-channels approach. 
The data for the $^{64}$Ni target above 0.1 mb are reproduced by including 
couplings to the low-lying $2^+$ and $3^-$ states and the mutual 
and two-phonon excitations of these states.
The calculations become more challenging as the fusing nuclei become 
softer and heavier, and excitations to multi-phonon states start to 
play an increasingly important role.
Thus it is necessary to include up to four-phonon excitations in order to 
reproduce the data for the $^{64}$Ni+$^{74}$Ge system. Similar calculations 
for $^{64}$Ni+$^{100}$Mo, and also for the symmetric $^{74}$Ge+$^{74}$Ge system, 
show large discrepancies with the data.
Possible ways to improve the calculations are discussed.
\end{abstract}
\maketitle

\section{Introduction}

Heavy-ion fusion data have in many cases been reproduced fairly well at 
energies close to the Coulomb barrier by coupled-channels calculations.
The best agreement is achieved for lighter or asymmetric systems, whereas 
the fusion of very heavy systems poses a serious challenge \cite{baha}.
Recently it was realized that the data for heavy systems 
are suppressed compared to single-channel calculations at energies 
above the Coulomb barrier \cite{newt1} and the suppression even 
persists in comparison to coupled-channels calculations \cite{newt2}.
There are also challenges in the theoretical description of heavy-ion 
fusion reactions at energies far below the Coulomb barrier, where
the measured cross sections are hindered compared to coupled-channels 
calculations \cite{syst}. While the suppression at high energy is 
largest for heavy systems, the hindrance of fusion far below the 
barrier appears to be a more general phenomenon. 

The structure input to coupled-channels calculations is commonly based 
on a vibrational or rotational model, with parameters determined from 
the lowest $2^+$ or $3^-$ excitations. These models make it possible
to calculate the coupling matrix elements to all orders in the
deformation amplitudes \cite{eslan,Takiful}, but they may not always provide 
a realistic representation of the actual structure of the reacting nuclei.
It is therefore desirable to be able to extract the necessary structure 
input from other measurements.  The most detailed information is available 
for quadrupole excitations up to very high spins, but there are too many 
states to consider in a practical coupled-channels calculation.  
One way to simplify the calculations is to construct effective multi-phonon 
states. This has been done for two-phonon quadrupole excitations \cite{alge}, 
and the procedure will be expanded here to the three-phonon level.

In view of the problems mentioned above it is of interest to make a 
systematic coupled-channels analysis of heavy-ion fusion reactions, 
ranging from lighter to the heavier systems, so that one can see exactly how 
the calculations fail to reproduce the data as the system gets heavier.
It is also important to know which and how many states are actually needed 
to make the calculations converge. For example, it is well known that  
couplings to high-lying states, like the giant resonances, produce an overall 
energy shift of the calculated fusion cross section, essentially without 
affecting the shape of the energy dependent cross section \cite{Takipol}.
If that is the case, then there is no need to include such high-lying states 
because their effect can be compensated by adjusting the radius or the depth 
of the ion-ion potential.

The basic coupled-channels calculations that will be presented  
include couplings to the one- and two-phonon excitations and to the mutual 
excitations of the lowest $2^+$ and $3^-$ states in projectile and target.
Such calculations reproduce the fusion cross sections rather well for lighter 
systems \cite{alge,eslan}.  The calculations will be expanded 
to include up to three- or four-phonon excitations, and they will be used
to analyze the data for the fusion of $^{64}$Ni with 
$^{64}$Ni \cite{nini}, $^{74}$Ge \cite{Beck}, and $^{100}$Mo \cite{nimo},
and also the fusion data for $^{74}$Ge+$^{74}$Ge \cite{Begege}.  
The projectiles and targets are fairly neutron rich in these systems so the 
couplings to transfer channels should have a modest effect on fusion.  

\section{Details of the calculations} 

The coupled-channels calculations are performed as described in Ref. 
\cite{alge}.  The main assumption is the rotating frame, - or isocentrifugal 
approximation. A basic feature of this approximation is that the magnetic 
quantum number $M$ is a conserved quantity. Thus if we start with two 
spin-zero nuclei in the entrance channel, then only the $M=0$ component 
of the excited states will be populated. That leads to a large reduction
in the number of channels one has to include, in comparison to a more
complete calculation.

The basic nuclear field that is required in the calculations described in
Ref. \cite{alge} is the ion-ion potential $U(r)$. The couplings to 
inelastic channels can be expressed in terms of radial derivatives of the 
ion-ion potential, and couplings up to second order in the deformation 
amplitudes $\alpha_{n\lambda\mu}$ will be included in the calculations. 
In the rotating frame approximation we only consider the $\mu=0$ component,
$\alpha_{n\lambda 0}$, so the nuclear interaction has form \cite{alge},
\begin{equation}
V(r,\alpha_{n\lambda 0}) = U(r) - 
\frac{dU(r)}{dr} \sum_{n\lambda} s_{n\lambda} + 
\frac{1}{2} \frac{d^2U(r)}{dr^2}
\Bigl[\bigl(\sum_{n\lambda} s_{n\lambda}\bigr)^2 -
\langle gs|\bigl(\sum_{n\lambda} s_{n\lambda}\bigr)^2 |gs\rangle\Bigr],
\end{equation} 
where $s_{n\lambda}=R_n\alpha_{n\lambda 0}\sqrt{(2\lambda+1)/(4\pi)}$ 
and $R_n$ is the nuclear radius.
The ion-ion potential $U(r)$ that will be used in the following is the 
empirical proximity-type potential developed in Ref. \cite{BW},
\begin{equation}
\label{BWP}
U(r) = - \frac{16\pi\gamma a R_1R_2}{R_1+R_2} \
\Bigl( 1 + \exp\Bigl[\frac{r-R_1-R_2-\Delta R}{a}\Bigr]\Bigr)^{-1}.
\end{equation}
Here the diffuseness $a$ depends on the mass numbers $A_n$ of the colliding 
nuclei, $1/a = 1.17[1+0.53(A_1^{-1/3}+A_2^{-1/3})]$ fm$^{-1}$,
and the nuclear radii are $R_n=1.2 A_n^{1/3}-0.09$ fm.
The parameter $\Delta R$ is zero in Ref. \cite{BW} but is introduced 
here as an adjustable parameter. Finally, the surface tension $\gamma$ 
is set to the constant value $\gamma$ = 0.95 MeV~fm$^{-2}$, whereas 
Ref. \cite{BW} included some isospin dependence in $\gamma$.

The fusion is simulated by ingoing wave boundary conditions at the radial
separation $r_{min}$, where the total potential in the elastic channel 
develops a minimum inside the Coulomb barrier. The nuclear potential in 
Eq. (1) is supplemented with a weak, short ranged, imaginary part $iW(r)$, 
where 
\begin{equation}
W(r) = \frac{-10 \ {\rm MeV}}{1+\exp([r-r_{min}]/a_W)},
\end{equation}
and $a_W$ is set to 0.2 fm. The effect of the imaginary potential
is to reduce certain oscillations in the calculated cross sections,
as discussed later on.

The one-phonon states that will be used are shown in Table I.
The coupling parameters are expressed in terms of the $\beta$-values, 
$\beta_\lambda^C$ and $\beta_\lambda^N$, for the Coulomb and nuclear 
fields, respectively. The Coulomb $\beta$-values are taken from the 
literature \cite{NNDC}.  The nuclear $\beta$-values are uncertain 
and they are often set equal to the Coulomb values. In this work a 
10\% higher nuclear $\beta$-value will be used. This is justified 
in the case of $^{64}$Ni from the analysis of inelastic scattering 
data \cite{Flem}, but such information is not always available.

In the rotating frame we only consider matrix elements of the
$\mu=0$ component of the deformation amplitudes $\alpha_{\lambda\mu}$
between $M=0$ magnetic sub-states \cite{alge}. The matrix 
elements are therefore of the form
\begin{equation}
\langle nI0| \alpha_{\lambda 0} | n'I'0 \rangle =
\langle I'0 \ \lambda 0|I0\rangle \
\frac{ \langle nI|| \alpha_{\lambda} || n'I'\rangle}{\sqrt{2I+1}},
\end{equation}
where $|nI0\rangle$ denotes a state at the $n$-phonon level with spin $I$.
The reduced matrix element determines the B-value,
\begin{equation}
B(E\lambda,nI\rightarrow n'I') = 
\Bigl(\frac{3ZeR_C^\lambda}{4\pi}\Bigr)^2 \
\frac{|\langle nI|| \alpha_{\lambda} || n'I'\rangle|^2}{2I+1},
\end{equation}
where $R_C=1.2A^{1/3}$ is the Coulomb radius of the nucleus being excited.
General matrix elements of the type given in Eq. (4) can be used to construct 
the couplings to effective two- and three-phonon states as discussed below 
for quadrupole excitations.

The coupling between the one- and two-phonon quadrupole states,
i.~e., between the $2^+_1$ states and the $0_2^+$, $2_2^+$ and $4_1^+$ 
states, can in some cases be obtained from the literature \cite{NNDC}.
Moreover, the two-phonon states are sometimes close in energy 
so it is convenient to combine them into one effective two-phonon 
state and thereby reduce the number of channels.
The procedure is described in Ref. \cite{alge}. 
The result (in the rotating frame approximation) is that the square of 
the coupling between the one-phonon and the effective two-phonon state is 
\begin{equation}
\label{ph2cou} 
\langle {\rm 2ph}| \alpha_{20} | {\rm 1ph}\rangle^2 =
\Bigl(\frac{4\pi}{3ZeR_C^2}\Bigr)^2
\sum_{I=0,2,4} \langle 2020|I0\rangle^2 \ B(E2, I\rightarrow 2),
\end{equation}
where the sum is over the spin $I$ of the three two-phonon states,
and $B(E2,I\rightarrow 2)$ are the reduced transition probabilities 
for the decay to the $2^+_1$ state.  The average two-phonon excitation 
energy is estimated by the energy weighted sum
\begin{equation}
E_{2{\rm ph}}= \sum_{I=0,2,4} E_2(I) \ 
\frac{|\langle {\rm 2ph},I0| \alpha_{20}| {\rm 1ph}, 20\rangle|^2}
{|\langle {\rm 2ph}| \alpha_{20} | {\rm 1ph}\rangle|^2}.
\end{equation}
This construction is possible for $^{74}$Ge and $^{100}$Mo as
shown in Table II and the results are included in Table I. 
The information about the two-phonon state in $^{64}$Ni is uncertain 
but it is consistent with a harmonic vibration, which is what is 
assumed in Table I.

In general, it is useful to characterize the coupling matrix elements
between the $n$-phonon and $(n-1)$-phonon states in terms of an
effective $\beta$ value, $\beta_\lambda(n{\rm ph})$, which is defined by
\begin{equation}
\langle n{\rm ph}| \alpha_{\lambda0} | (n-1){\rm ph}\rangle =
\sqrt{n} \ \frac{\beta_\lambda(n{\rm ph})}{\sqrt{2\lambda+1}}.
\end{equation} 
The advantage of this representation is that the effective $\beta$-values
are identical in the harmonic oscillator model. Consistent with this definition
one can also define the effective B-value,
\begin{equation}
B(E\lambda,n{\rm ph}\rightarrow (n-1){\rm ph}) = 
\Bigl(\frac{3ZeR_C^\lambda}{4\pi}\Bigr)^2 \ 
\frac{n \ \beta_\lambda^2(n{\rm ph})}{2\lambda+1}.
\end{equation} 

When the structure of multi-phonon excitations deviates from the
harmonic oscillator model, it can become difficult to calculate matrix 
elements of a general nuclear interaction.  However, the calculation
is not that difficult if the interaction only contains terms that
are linear and quadratic in the deformation amplitudes, as in Eq. (1).
Matrix elements of the quadratic terms can be calculated by 
inserting a complete set $|n'{\rm ph}\rangle$ of intermediate states,
\begin{equation}
\langle n_2{\rm ph}| \alpha_{\lambda0}^2 | n_1{\rm ph}\rangle =
\sum_{n'} \
\langle n_2{\rm ph}| \alpha_{\lambda0} | n'{\rm ph}\rangle \
\langle n'{\rm ph}| \alpha_{\lambda0} | n_1{\rm ph}\rangle.
\end{equation}
The intermediate couplings in this expression that are not known will be 
estimated in the harmonic limit by extrapolation from nearby known states.

The information about octupole excitations is usually limited to the 
one-phonon excitation.  The octupole modes shown in Table I have 
been modeled as harmonic vibrations.
The basic two-phonon calculations, referred to as the PH-2 calculations,
will include all of the one- and two-phonon states shown in Table I, in 
addition to the mutual excitations of the one-phonon states. 
All calculations are furthermore restricted by a 7 MeV cutoff in 
excitation energy.  

\section{Systematics of comparison with data}

The measured fusion cross sections for $^{64}$Ni on the three targets: 
$^{64}$Ni \cite{nini}, $^{74}$Ge \cite{Beck}, and $^{100}$Mo \cite{nimo}, 
are compared to two sets of calculations in Fig. 1. The solid curves show
the basic two-phonon coupled-channels calculations (PH-2) described in 
the previous section, and the dashed curves are the results obtained in 
the no-coupling limit using the same ion-ion potential.
There is an essentially constant energy shift between the solid and dashed 
curves at low energies, and the energy shift is seen to increase as the 
target gets heavier and softer.

All calculations shown in Fig. 1 were based on the value $\Delta R$ = 0.1 fm 
in the ion-ion potential, Eq. (2).  This value was chosen because it provides 
the best fit to the $^{64}$Ni+$^{64}$Ni fusion data that are larger than 0.1 mb.
The best fit to all of the $^{64}$Ni+$^{64}$Ni data is actually quite poor,
with a $\chi^2/N$=10.  The reason is that the coupled-channels calculations 
cannot reproduce the fusion hindrance that occurs far below the Coulomb barrier 
\cite{nini}, so it is better to exclude that energy region from the analysis. 

The $\chi^2$ values per point for different calculations are summarized 
in Table III.  By comparing the values obtained in the one-phonon (PH-1) 
and two-phonon (PH-2) calculations it is seen that the mutual and two-phonon 
excitations play a very important role in improving the fit to the data. 
The improvement is mainly caused by the second-order (or quadratic) term 
in the nuclear interaction, Eq. (1). This has been demonstrated previously 
in Refs. \cite{eslan,Takiful}.
It is also seen that the PH-2 fit to the $^{64}$Ni+$^{74}$Ge data is not 
as good as the fit to the $^{64}$Ni+$^{64}$Ni data, and the fit to the 
$^{64}$Ni+$^{100}$Mo data is even worse. In the next section we shall
try to improve the fits by including more channels in the calculations.  

The third column shows the overall energy shift $\Delta E$ that is 
required to minimize the $\chi^2$ for a given calculation and data set.
It is seen that the shift depends on the channels that are included 
in the calculations.  The energy shift is effectively equivalent to 
changing the radius parameter $\Delta R$ of the ion-ion potential, 
Eq. (\ref{BWP}).
Thus one should realize that one cannot easily beforehand predict 
the best value of $\Delta R$, because the value that gives the best 
fit to the data depends on channels that are included in the calculation. 
This feature was recognized in Ref. \cite{Takipol} in a study of
the influence of couplings to high-lying states on fusion.
Another example is the two-phonon $3^-$ state in $^{64}$Ni, which was 
excluded from the calculations by the 7 MeV cutoff in excitation energy.
If this state and also the mutual excitations of the $3^-$ states in projectile
and target are included in the analysis of the $^{64}$Ni+$^{64}$Ni
fusion data above 0.1 mb, one obtains the same good fit as shown
in Table III but the required energy shift is now $\Delta E$ = 0.27 MeV. 
The same fit is achieved by reducing $\Delta R$ from 0.10 fm to 0.07 fm. 

\subsection{Logarithmic derivative}

One way to illustrate the behavior of the fusion cross section at low 
center-of-mass energies, $E_{c.m.}$, is to plot the logarithmic derivative 
of the energy weighted fusion cross section, which is defined as \cite{syst}
\begin{equation}
\label{logderiv}
L(E_{c.m.}) = \frac{1}{E_{c.m.}\sigma_f} \ 
\frac{d(E_{c.m.}\sigma_f)}{dE_{c.m.}}.
\end{equation}
The results for the three heavy-ion systems are shown in Fig. 2. 
The experimental values are seen to increase steeply with decreasing energy. 
The top lines in Fig. 2 show the logarithmic derivative for constant $S$ 
factor \cite{syst}, i.~e. $L_{CS}=\pi\eta/E_{c.m.}$, where $\eta$ is the 
Sommerfeld parameter. It is seen that the data for the $^{64}$Ni and 
$^{100}$Mo targets intersect the constant $S$ factor curves. 
The experimental $S$ factor, $S=E_{c.m.}\sigma_f\exp(2\pi\eta)$, will 
therefore exhibit a maximum, and the energy where that occurs has been used to 
characterize the onset of the low-energy fusion hindrance \cite{nini,nimo}. 
The data for the $^{74}$Ge target have not reached the constant 
$S$ factor limit but it is likely they will if measurements were performed 
at lower energies.

The logarithmic derivatives obtained from the no-coupling calculations 
rise steeply near the Coulomb barrier but level off at lower energies.  
The results for the PH-2 coupled-channels calculations 
show a similar behavior. They are just shifted to lower energies.
They exhibit some oscillations at the lowest energies. The magnitude 
of the oscillations depend on where the ingoing wave boundary conditions 
are imposed and on the strength of the imaginary optical potential.
However, the oscillations are not essential since they have not been
observed experimentally. The important point is that it has not yet 
been possible, within the coupled-channels approach, to develop a 
credible description that reproduces the data at the lowest energies.

\subsection{High-energy behavior}

The cross sections shown in Fig. 1 are plotted on a linear scale 
in Fig. 3. It is unfortunate that the measurements  do not reach 
larger cross sections, so it is difficult to assess the suppression 
of the data compared to the no-coupling limit, which was identified 
in Ref. \cite{newt1} for cross sections larger than 200 mb.
It is seen, however, that the coupled-channels calculations agree 
fairly well with the data points that are above the 200 mb limit,
and they are shifted to higher energies (i.e., they are suppressed) 
when compared to the no-coupling calculations. 
The suppression is mainly caused by the long-range Coulomb excitation
of the low-lying quadrupole states, and it is largest for the $^{74}$Ge 
and $^{100}$Mo targets, consistent with the fact that the quadrupole 
mode is particularly soft in these two nuclei.

Another way to illustrate the behavior of the fusion cross sections 
for the three systems is to plot the derivative of $E_{c.m.}\sigma_f$, 
which is shown in Fig. 4.  The behavior at energies far above the Coulomb 
barrier, $V_{CB}$, is often parametrized as 
\begin{equation}
\sigma_f=\pi R_f^2 \ \Bigl(1-\frac{V_{CB}}{E_{c.m.}}\Bigr).
\end{equation}
The derivative of $E_{c.m.}\sigma_f$ should therefore approach the constant 
value $\pi R_f^2$ at high energy. The curves shown in Fig. 4 do approach 
a constant value at high energies and the data are also consistent with that 
behavior. There is some uncertainty in the highest data point for the 
$^{64}$Ni+$^{100}$Mo system because the estimated contribution from 
fission is large. It is also seen that the data are suppressed at energies 
near the Coulomb barrier and enhanced at lower energies compared to the 
PH-2 calculation. There are similar but more modest discrepancies for 
the $^{64}$Ni+$^{74}$Ge system.

\section{Effects of multi-phonon excitations}

A simple way to expand the PH-2 calculations presented in the
previous section is to include all mutual excitations of the
states given in Table I up to three-phonon (2PH-3) or four-phonon
excitations (2PH-4). The results of the $\chi^2$ analysis using such 
calculations, again with a maximum excitation energy cutoff of 7 MeV, 
are shown in Table III for the $^{74}$Ge and $^{100}$Mo targets, and also 
for the $^{74}$Ge+$^{74}$Ge systems, which will be discussed below.
It is seen that the expanded calculations give a much better fit to the 
$^{64}$Ni+$^{74}$Ge data, whereas the improvements for the heavier 
systems are modest. 

In addition to the mutual excitations discussed above, one can
also explicitly include three- and four-phonon excitations in
the calculations. The problem is that the energy and transition 
strengths of such states are poorly known. The best known transitions
in heavy nuclei are the quadrupole transitions and below we estimate
the effect on fusion of couplings to an effective three-phonon 
quadrupole state.  There are many states at the three-phonon level 
but it is possible to lump them together into one effective 
three-phonon state as described in the Appendix.

\subsection{Calculations for $^{64}$Ni+$^{74}$Ge}

The quadrupole mode in $^{74}$Ge is rather soft and there 
are several even parity states at the three-phonon level.
Unfortunately, the knowledge about the couplings to these states 
is poor so it is not possible to construct an effective three-phonon
state using the method described in the Appendix.
It is therefore assumed in the following that the three-phonon 
quadrupole state is at 2 MeV and that the associated 
$\beta$-values are the same as for the two-phonon state, i.~e.,
$\beta_2^C$(3ph)=0.217 and $\beta_2^N$(3ph)=0.239.

The calculations that include the three-phonon quadrupole mode, 
in addition to all of the mutual excitations of the states shown 
in Table I up to the four-phonon level, are denoted by 3PH-4. 
These calculations are again restricted to excitation energies 
below a 7 MeV cutoff and consist of 35 coupled channels.  
They can be compared to the 2PH-4 calculations, which do not
include the three-phonon quadrupole mode in $^{74}$Ge. 
From table III it is concluded that the three-phonon state is not 
very important since both calculations require the same energy 
shift to fit the data and give essentially the same $\chi^2$.
The required energy shift of -0.57 MeV is equivalent 
to increasing the $\Delta R$ parameter from 0.10 to 0.16 fm.

The calculated 3PH-4 cross sections are shown by the solid curve in 
the top part of Fig. 5. The results of the PH-2 and 
PH-1 calculations and the no-coupling limit are also shown for 
comparison.  All calculations shown here were based on the parameter 
value $\Delta R$=0.16 fm, which minimizes the $\chi^2$ fit to the 
data in the 3PH-4 calculations. It is seen that the 3PH-4 and
the PH-2 calculations do not differ much, except at the lowest energies.
In other words, multi-phonon excitations (beyond PH-2) do not play
a very dramatic role in the calculation of the $^{64}$Ni+$^{74}$Ge
fusion cross section; they just provide a fine-tuning that produces 
a better $\chi^2$ fit to the data.

The logarithmic derivatives of the cross sections are
shown in the bottom part of Fig. 5. 
The shape of the calculated $L(E_{c.m.})$ clearly improves
in comparison with the low-energy data as more channels are 
included in the calculations. 
Some discrepancy is beginning to develop at the lowest energy 
where the calculated $L(E_{c.m.})$ saturates, whereas 
the measured values keep growing with decreasing energy.
This behavior is consistent with the systematic trend of the 
fusion hindrance phenomenon at extreme subbarrier \cite{nimo}, 
which was observed in Fig. 2 for the other two systems.

\subsection{Calculations for $^{64}$Ni+$^{100}$Mo}

The experimental structure information about the states associated with the
three-phonon quadrupole excitation in $^{100}$Mo is shown in Table IV.
The excitation energies and most of the couplings to two-phonon states 
are known, and the two that are not known can be estimated in the 
harmonic oscillator model (the values in parenthesis). 
In that model one has the sum rule
\begin{equation}
\sum_{I_2} B(E2,I_3\rightarrow I_2) = 3 \ B(E2, 2\rightarrow 0),
\end{equation}
i.~e., the sum is three times the B-value for the one-phonon 
excitation.  That would give a sum of 112 W.u., according to Table I. 
It is seen that the sums in Table IV for different values of the spin 
$I_3$ are less than the harmonic limit. 

We can now use the expressions derived in the Appendix to estimate
the properties of the effective three-phonon state. The expression
(\ref{alf3}) is based on amplitudes, whereas the B-values are proportional 
to the square of these amplitudes. To proceed we assume that the 
amplitudes are positive and determined by the square root of the 
B-values. This assumption is correct for the harmonic oscillator.
If we use all of the transition strengths shown in Table IV,
including the two harmonic estimates of the two unknown strengths
(indicated by the `?'), we obtain $\beta_2^C({\rm 3ph})$ = 0.182 
and a three-phonon excitation energy of 1.65 MeV.  
The nuclear coupling is set 10\% higher, i.~e. $\beta_2^N$(3ph)=0.2.
These are the values that will be used in the 3PH-3 and 3PH-4
calculations discussed below. 

The estimated three-phonon coupling strength is weaker than the 
harmonic limit (which gives $\beta_2^C=0.231$) but the excitation 
energy is not much different from the harmonic value (1.61 MeV). 
Another limit is to include only the known transition strengths 
in Table IV, and exclude the harmonic estimates of the
two unknown strengths. That gives the value $\beta_2^C({\rm 3ph})$ 
= 0.161 and an excitation energy of 1.70 MeV. 
This shows that the three-phonon excitation energy is quite 
accurately determined, whereas the $\beta$-value may have 
an uncertainty of 10\%,

The results of the $\chi^2$ analysis,
which was based on the radius parameter value $\Delta R$ = 0.10 fm,
are shown in Table III. The best fit is obtained with the 3PH-3 calculation 
but the fit is still poor. When the analysis is restricted to cross sections 
that are larger than 0.1 mb, the $\chi^2/N$ is reduced from 25 to 15.
The poor fit to the data is therefore not entirely due to the fusion 
hindrance phenomenon, which occurs at smaller cross sections.

Inspecting the fusion data for $^{64}$Ni+$^{100}$Mo shown in Fig. 1 
it appears that the fusion hindrance sets in at cross sections that are
smaller than 1 $\mu$b.  To fit the data above 1 $\mu$b with the 3PH-3 
calculation requires an energy shift of 0.7 MeV, or a new radius
parameter of $\Delta R$ = 0.04 fm. To make a reasonable comparison 
with the data, all calculations shown in Fig. 6 have therefore been 
based on this value for the radius parameter.

In the top panel of Fig. 6 it is seen that the 3PH-3 calculation agrees 
with the largest cross section and the 0.92 mb cross section measured 
at 127.5 MeV, but it is too high in between.
The good agreement at the highest energy could be misleading because
50\% of the fusion cross section that is shown (solid point) is an
estimated contribution from fission. The measured evaporation residue
cross sections are shown by the open circles. If the correction for 
fission were smaller, then there would be a general suppression of 
the data at energies above the Coulomb barrier, in agreement with 
systematics \cite{newt1,newt2}.  

The discrepancy between calculations and measurements at low 
energies is emphasized by comparing the logarithmic derivatives, 
which are shown in the bottom part of Fig. 6.
The discrepancy is large, and the calculated curves fail to
reproduce the value at the lowest energy, where the fusion hindrance 
phenomenon sets in.  Let us finally show some results for $^{74}$Ge+$^{74}$Ge 
in order see how the discrepancies evolve with mass asymmetry.

\subsection{Results for $^{74}$Ge+$^{74}$Ge}

The results of calculations for the $^{74}$Ge+$^{74}$Ge system,
using the radius parameter value $\Delta R$ = 0.10 fm, are compared 
with the data \cite{Begege} in Fig. 7.
The data are the measured evaporation residue cross sections
whereas the fission cross sections were estimated to be small,
about 15 mb at the highest energy.
The three-phonon quadrupole state used in the 3PH-4 calculation is
the same as used in Sect. IV.A, whereas the 2PH-4 calculation
does not include that state. By comparing the two calculations
it is seen that couplings to the three-phonon quadrupole state have 
a significant influence and improves the shape of the calculated 
cross section in comparison with the data. Thus it reduces the 
cross section at higher energies and enhances it at energies below 
the Coulomb barrier. This is what is needed to improve the fit
to the data but the discrepancy with the data remains large, 
with a $\chi^2/N$=17. The demonstrated sensitivity to the three-phonon 
quadrupole state indicates that the calculation has not converged.  

The logarithmic derivatives are shown in the bottom part of Fig. 7.
The two coupled-channels calculations exhibit oscillations 
that are out of phase at low energy and the 2PH-4 calculation 
has a bump near 120 MeV. These differences should not be taken too
seriously because the calculations have not converged with respect 
to multi-phonon excitations, as explained above. 
The discrepancies with the data in Fig. 7 are similar to 
those seen in Fig. 6 for the fusion of $^{64}$Ni+$^{100}$Mo. 
The measured fusion cross sections are suppressed at high energies
compared to the calculations, in agreement with systematics 
\cite{newt1,newt2}, and they are enhanced below the Coulomb barrier, 
where they fall off at a slower pace when compared to the calculations. 

\subsection{Further improvements}

In order to improve the calculations of the fusion of heavy, soft nuclei, 
one would have to include excitations to higher multi-phonon states. 
That would create several problems. One problem is that the structure 
of such states is often poorly known.
Another problem is that the nuclear interaction, Eq. (1), only 
includes terms up to second order in the deformation amplitudes, 
whereas matrix elements associated with multi-phonon states 
would be sensitive to higher-order terms.
In the harmonic oscillator model one can calculate the nuclear 
coupling matrix elements to all orders in the deformation 
amplitudes \cite{eslan,Takiful}. Unfortunately, the connection 
to the actual structure of the reacting nuclei may be poor in such 
calculations. The most practical solution to these problems would 
be to develop and apply more realistic structure models, such as 
those presented in Ref. \cite{Takianh}.

\section{Conclusions}

The calculated fusion cross sections that were presented and compared
to measurements illustrate nicely how multi-phonon excitations play an 
increasingly important role as the fusing nuclei become softer and heavier. 
The basic two-phonon calculations reproduce quite well the data for 
the lightest system $^{64}$Ni+$^{64}$Ni (except at extreme subbarrier 
energies) but it is necessary to include up to three- or four-phonon 
excitations in order to reproduce the $^{64}$Ni+$^{74}$Ge fusion data. 
In the analysis of the data for the two heavy systems, 
$^{64}$Ni+$^{100}$Mo and $^{74}$Ge+$^{74}$Ge, it was not possible 
to achieve a good fit by including up to four-phonon excitations.  
The subbarrier fusion data were enhanced and the data above the Coulomb 
barrier were suppressed compared to the most complete calculations. 
It is difficult to ascertain which of these two discrepancies is the most 
dominant because the radius parameter of the ion-ion potential, which 
determines the energy scale of the calculated fusion cross sections, 
cannot be predicted accurately.

The observed suppression of the data above the Coulomb barrier compared 
to the no-coupling limit is consistent with the results of the systematic 
studies by Newton et al. \cite{newt1,newt2}. The fusion cross sections 
obtained in the coupled-channels calculations are also suppressed 
compared to the no-coupling limit.
This is caused by the long-range Coulomb excitation
which pushes the surfaces of the reacting nuclei away from each other 
whereby the attractive nuclear force is reduced.  However, the calculated 
suppression is apparently not large enough to explain the data. 
It has been suggested that the remaining discrepancy could be due to 
deep inelastic reactions \cite{newt2}.

The enhancement of the data at energies below the Coulomb barrier
(ignoring for a moment the hindrance phenomenon at extreme subbarrier energies) 
indicates that additional channels or couplings need to be included
in the coupled-channels calculations. 
One possibility is transfer reactions but they should play a minor 
role in the fusion of the symmetric $^{74}$Ge+$^{74}$Ge system. 
The most natural explanation is the limitations of the nuclear 
interaction which was expanded up to second order in the deformation 
parameters. This approximation works very well in calculations 
of fusion cross sections for lighter and stiff systems,
but it becomes unrealistic for heavy and soft systems, 
where the calculations become sensitive to multi-phonon excitations 
and therefore to the nuclear interaction at large deformation amplitudes
\cite{Takiful}.

A serious complication and uncertainty in the calculation of fusion 
cross sections for heavy and soft nuclei is the sensitivity to poorly 
known multi-phonon states.  
One example is the two-phonon octupole state. Fortunately, this 
state did not play a very crucial role in the calculations that 
were presented for $^{64}$Ni+$^{64}$Ni. 
The main effect of couplings to this state was an overall shift in 
energy, whereas the shape of the cross section was not affected. 
The knowledge of quadrupole excitations is much better.
Thus it was demonstrated that it is possible to construct effective 
two- and three-phonon states in $^{100}$Mo from detailed structure 
information.  Such a construction may become more difficult 
at the four-phonon level.  

In addition to improving the structure input for multi-phonon excitations
and the nuclear interaction at large deformations, one should also seek 
an explanation for what causes the fusion hindrance in measurements at 
extreme subbarrier energies. That could also be related to the 
parametrization of the ion-ion potential, as suggested by Dasso 
and Pollarolo \cite{Dasso}.

\begin{acknowledgments} 
This work was supported by the U.S. Department of Energy, 
Office of Nuclear Physics, under Contract No. W-31-109-ENG-38.
\end{acknowledgments}

\appendix*
\section{Two- and three-phonon excitations}

The number of states at the two- and three-phonon level of
quadrupole excitations is so large that it is convenient to lump 
them together into effective two- and three-phonon states. 
The procedure for doing that in the rotating frame approximation was 
discussed in Ref. \cite{alge} for two-phonon excitations and below 
it is shown how it can be generalized to three-phonon level.

Let us include all couplings that are linear and quadratic in the 
deformation amplitudes and denote the associated radial form
factors by $F_1(r)$ and $F_2(r)$. 
The radial wave functions for the $n$-phonon state with spin $I$ 
is denoted by $\psi_n(I)$, and the coupling of the $(n,I)$ state 
to the $(n-1,I_{n-1})$ state is
\begin{equation}
\alpha_n(I_n,I_{n-1}) =
\langle I_{n-1} 0 \ 20| I_n0\rangle
\frac{\langle nI||\alpha_2|| (n-1)I_{n-1}\rangle} {\sqrt{2I_n+1}},
\end{equation}
according to Eq. (4).  The notation can be simplified for the one- 
and two-phonon couplings by defining $\alpha_1=\alpha_1(2,0)$ 
and $\alpha_2(I)=\alpha_2(I,2)$.
With these definitions, the coupled equations have the following 
form in the rotating frame approximation 
\begin{equation}
\label{cc0}
(H-E) \psi_0 = - F_1 \ \alpha_1 \ \psi_1 
- F_2 \sum_I \alpha_1 \alpha_2(I) \ \psi_2(I),
\end{equation}
\begin{equation}
\label{cc1}
(H - E_1) \psi_1 = - F_1 \alpha_1 \ \psi_0 - 
F_1 \sum_{I_2} \alpha_2(I_2) \ \psi_2(I_2) - F_2
\sum_{I_2,I_3} \alpha_2(I_2) \alpha_3(I_3,I_2) \ \psi_3(I_3),
\end{equation}
\begin{equation}
\label{cc2}
(H - E_2(I_2)) \psi_2(I_2) =  
- F_2 \alpha_2(I_2) \alpha_1 \ \psi_0 
- F_1 \alpha_2(I_2) \ \psi_1 
- F_1 \sum_{I_3} \alpha_3(I_3,I_2) \ \psi_3(I_3),
\end{equation}
\begin{equation}
\label{cc3}
(H - E_3(I_3)) \psi_3(I_3) = 
- F_1 \sum_{I_2} \alpha_3(I_3,I_2) \ \psi_2(I_2)
- F_2 \sum_{I_2} \alpha_3(I_3,I_2) \alpha_2(I_2) \ \psi_1,
\end{equation}
where $H$ is the scattering Hamiltonian in the rotating frame,
which is assumed to be the same for all channels. 
The available energy in the channel $(n,I)$ is denoted by 
$E_n(I)=E-\epsilon_n(I)$, where $\epsilon_n(I)$ is the excitation energy. 

The structure of the above equations is actually quite simple.
An effective two-phonon state, with coupling strength $\alpha_2$
and scattering wave function $\psi_2$, can be introduced by 
the substitution
\begin{equation}
\label{ph2}
\alpha_2 \psi_2 =  \sum_I \alpha_2(I) \ \psi_2(I), 
 \ \ {\rm where} \ \ \ \alpha_2^2 = \sum_I \alpha_2(I)^2.    
\end{equation}
A similar substitution can be made for the three-phonon state,
\begin{equation}
\label{ph3}
\alpha_2\alpha_3 \ \psi_3 = 
\sum_{I_2,I_3} 
\alpha_2(I_2) \alpha_3(I_3,I_2) \ \psi_3(I_3).
\end{equation}
where the three-phonon coupling strength $\alpha_3$ is determined by
\begin{equation}
\label{alf3}
(\alpha_2\alpha_3)^2 = \sum_{I_3} 
\Bigl[\sum_{I_2} \alpha_2(I_2) \alpha_3(I_3,I_2)\Bigr]^2.
\end{equation}
These substitutions simplifies the coupled equations 
(\ref{cc0}-\ref{cc3}).
The first two become
\begin{equation}
\label{cef0}
(H-E) \psi_0 = - F_1 \ \alpha_1 \ \psi_1 
- F_2 \alpha_1 \alpha_2 \ \psi_2,
\end{equation}
\begin{equation}
\label{cef1}
(H - E_1) \psi_1 = - F_1 \alpha_1 \ \psi_0 - 
F_1 \alpha_2 \ \psi_2 - F_2
\alpha_2\alpha_3 \ \psi_3,
\end{equation}
which is just the form we want for a zero- and one-phonon state that 
are coupled to each other and to the two- and three-phonon states.

The Eq. (\ref{cc2}) can be brought into a similar
form if we assume that the two-phonon energies
are almost degenerate, $E_2(I)\approx E_2$.
We can then multiply Eq. (\ref{cc2}) by $\alpha_2^{-1}\alpha_2(I_2)$,
sum over $I_2$, and use the substitutions (\ref{ph2}) and (\ref{ph3}) 
to obtain
\begin{equation}
\label{cef2}
(H - E_2) \psi_2 =  
- F_2 \alpha_2 \alpha_1 \ \psi_0 
- F_1 \alpha_2 \ \psi_1 
- F_1 \alpha_3 \ \psi_3.
\end{equation}

The Eq. (\ref{cc3}) can be dealt with in a similar way assuming a
near degeneracy of the three-phonon states, $E_3(I)\approx E_3$.
Thus we can multiply Eq. (\ref{cc3}) by
$(\alpha_2\alpha_3)^{-1}\alpha_2(I_2)\alpha_3(I_3,I_2)$, sum over 
$I_2$ and $I_3$, and use Eqs. (\ref{ph3}) and (\ref{alf3}) to obtain
\begin{equation}
\label{cef3}
(H - E_3) \psi_3 = 
- F_1 \alpha_3 \ \psi_2
- F_2 \alpha_3\alpha_2 \ \psi_1.
\end{equation}
Actually, to derive the first term on the right hand side of Eq. 
(\ref{cef3}) one also needs the assumption 
$\psi_2(I_2)$ = $\alpha_2^{-1}\alpha_2(I_2) \psi_2$, which 
is not unreasonable because it is consistent with Eq. (\ref{ph2}) 
and it is exact for $\alpha_3=0$, as can be seen from Eqs. (\ref{cc2}) 
and (\ref{cef2}).
That completes the derivation.  If the states at the two- or
three-phonon level are not quite degenerate one can construct the 
effective two- and three-phonon excitation energies from the energy 
weighted sums
\begin{equation}
\epsilon_2 = \frac{1}{\alpha_2^2} \
\sum_I |\alpha_2(I)|^2 \ \epsilon_2(I),
\end{equation}
\begin{equation}
\epsilon_3 = \frac{1}{(\alpha_2\alpha_3)^2}  
\sum_{I_3} \Bigl[\sum_{I_2} \alpha_2(I_2) \alpha_3(I_3,I_2)\Bigr]^2 \ 
\epsilon_3(I_3).
\end{equation}

\begin{table}
\caption{Excitation energies and coupling strengths used in the calculations.
The values for $^{64}$Ni are the same as used in \cite{nini}.
The B-values for $^{74}$Ge are from Refs. \cite{Lecomte,Spear} and from
Ref. \cite{NNDC} for $^{100}$Mo.
The effective two-phonon quadrupole states are determined in Table II,
and the two-phonon octupole states are estimated in the harmonic 
oscillator model.}
\begin{ruledtabular}
\begin{tabular} {|c|c|c|c|c|c|}
Nucleus &
$\lambda^\pi$ &  $E_x$ (MeV) & 
B(E$\lambda$) (W.u.) & \ $\beta_\lambda^C$ & $\beta_\lambda^N$ \\
\colrule
$^{64}$Ni  & $2^+$   & 1.346 & 8.6    & 0.165 & 0.185  \\
        & 2ph($2^+$) & 2.692 & (17.2) & 0.165 & 0.185  \\
 Z=28      & $3^-$   & 3.560 & 12     & 0.193 & 0.200  \\
\colrule
$^{74}$Ge  & $2^+$   & 0.596 &   33 & 0.285 & 0.314  \\
        & 2ph($2^+$) & 1.362 &   38 & 0.217 & 0.239  \\
  Z=32     & $3^-$   & 2.536 &  8.8 & 0.145 & 0.160  \\
        & 2ph($3^-$) & 5.072 & (17.6) & 0.145 & 0.160 \\
\colrule
$^{100}$Mo & $2^+$   & 0.536 &  37.4 & 0.231 & 0.254  \\
        & 2ph($2^+$) & 1.002 &  68  & 0.222 & 0.244  \\ 
  Z=42     & $3^-$   & 1.908 &  35  & 0.220 & 0.242  \\
        & 2ph($3^-$) & 3.816 & (70) & 0.220 & 0.242  \\
\end{tabular}
\end{ruledtabular}
\end{table}

\begin{table}
\caption{Energies and reduced transition probabilities (in W.u.) of the $0^+$, 
$2^+$, and $4^+$ states associated with a two-phonon quadrupole excitation. 
The values for 
$^{100}$Mo are from \cite{NNDC}, and the
values for $^{74}$Ge are from \cite{Lecomte}. The last two columns shows
the energy and coupling strength of the effective two-phonon state.}
\begin{ruledtabular}
\begin{tabular} {|c|c|c|c|c|c|c|}
Nucleus &  States:   & $0_2^+$ & $2_2^+$ & $4_1^+$ & 
Eff. 2ph & $\beta_2$(2ph) \\ 
\colrule
$^{74}$Ge & $E_I$ (MeV):        & 1.483 & 1.204 & 1.464 & 1.362 &  -    \\
      & B(E2,I$\rightarrow 2$): & $<$22  &  54.2 & 36.1  & $<$38.5 & $<$0.217  \\
\colrule
$^{100}$Mo & $E_I$ (MeV):           & 0.695 & 1.064 & 1.136 & 1.002 &  -    \\
           & B(E2,I$\rightarrow$2$^+$): &   92  &   51  &   69  &  68   & 0.222  \\
\end{tabular}
\end{ruledtabular}
\end{table}

\begin{table}
\caption{Analysis of fusion data for different heavy-ion systems
using the nuclear interaction, Eq. (1) and (2), with $\Delta R$=0.10 fm,
and a 7.0 MeV cutoff in excitation energy. } 
\begin{ruledtabular}
\begin{tabular} {|c|c|c|c|c|}
System  &  Calculation & $\Delta E$ (MeV) & $\chi^2/N$ & Data Ref. \\
\colrule
$^{64}$Ni+$^{64}$Ni & PH-2 &  0.9 &  10  & all data \cite{nini} \\
                    & PH-2 & 0.03 &  0.5 & $\sigma_f>$0.1 mb \\
\colrule
$^{64}$Ni+$^{74}$Ge  & PH-1  & -1.53  & 15.6 & all data \cite{Beck} \\
           & PH-2  & -0.62  & 6.4  &    \\
           & 2PH-3 & -0.56  & 3.9  &    \\
           & 2PH-4 & -0.62  & 3.1  &    \\
           & 3PH-4 & -0.57  & 2.6  &    \\
\colrule
$^{64}$Ni+$^{100}$Mo & PH-1   &  0.18 & 44   & all data \cite{nimo} \\
           & PH-2   &  0.   & 33   &          \\
           & 2PH-3  &  0.35 & 27   &          \\
           & 3PH-3  &  0.61 & 25   &          \\
           & 2PH-4  &  0.13 & 31   &          \\
           & 3PH-4  &  0.42 & 34   &          \\
\colrule
$^{64}$Ni+$^{100}$Mo & PH-1  &  0.20  &  38   & $\sigma_f>0.1$ mb \\
           & PH-2  &  0.30  &  24   & \cite{nimo} \\
           & 2PH-3 &  0.70  &  18   &              \\
           & 3PH-3 &  0.94  &  15   &              \\
           & 2PH-4 &  0.56  &  25   &              \\
           & 3PH-4 &  0.85  &  27   &              \\
\colrule
$^{74}$Ge+$^{74}$Ge & PH-1  & +4.50  & 30  & all data \cite{Begege} \\
                    & PH-2  & -0.50  & 25  &     \\
                    & 2PH-4 & +0.12  & 26  &     \\
                    & 3PH-4 & +0.20  & 17  &     \\
\end{tabular}
\end{ruledtabular}
\end{table}

\begin{table}
\caption{Three-phonon quadrupole excitation in $^{100}$Mo. 
The B-values are shown (in W.u.) for the known E2 transitions, 
from the ($I_3$ = 0, 2, 4, 6) three-phonon states to the
($I_2$ = 0, 2, 4) two-phonon states \cite{NNDC}.
Unknown values are indicated by '?'.
The values in parenthesis were obtained for a harmonic vibration.
The second last row shows the sum of the B-values for each $I_3$ spin state.
The last row shows the excitation energies of the states associated 
with the three-phonon excitation.}
\begin{ruledtabular}
\begin{tabular} {|c|c|c|c|c|}
$B(E2,I_3\rightarrow I_2)$ &  $I_3$ =0  &  2  &  4   &   6   \\
\colrule
 $I_2$ = 0 &   ---   &  14 (52) &    ---    & ---  \\
    2      & ? (112) &  ?  (21) &  30 (59)  &  --- \\
    4      &   ---   &  36 (38) &  28 (53)  &  94 (112) \\
\colrule
  Sum =    & ? (112) & $>$ 50 (112) &  58 (112) &  94 (112) \\
\colrule
$E_x(I_3)$ (MeV) & 1.5046 & 1.4639 & 1.7715 & 1.8469 \\
\end{tabular}
\end{ruledtabular}
\end{table}


\begin{figure}
\includegraphics[width = 12cm]{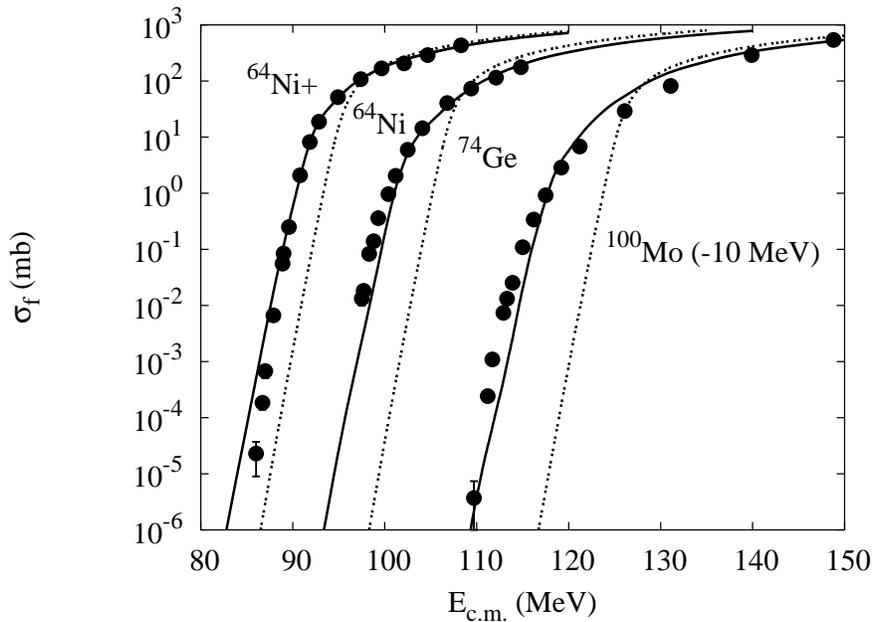}
\caption{\label{alf} Fusion cross sections for the three systems
$^{64}$Ni+$^{64}$Ni \cite{nini}, $^{64}$Ni+$^{74}$Ge \cite{Beck}, and
$^{64}$Ni+$^{100}$Mo \cite{nimo} as functions of the center-of-mass
energy $E_{c.m.}$. Note that the results for $^{64}$Ni+$^{100}$Mo have 
been shifted by -10 MeV.
The dashed curves show the no-coupling limit, and the solid curves are 
the results of the PH-2 coupled-channels calculations described in the text.
All calculations were based on the radius parameter $\Delta R$ = 0.10 fm.}
\end{figure}

\begin{figure}
\includegraphics[width =12cm]{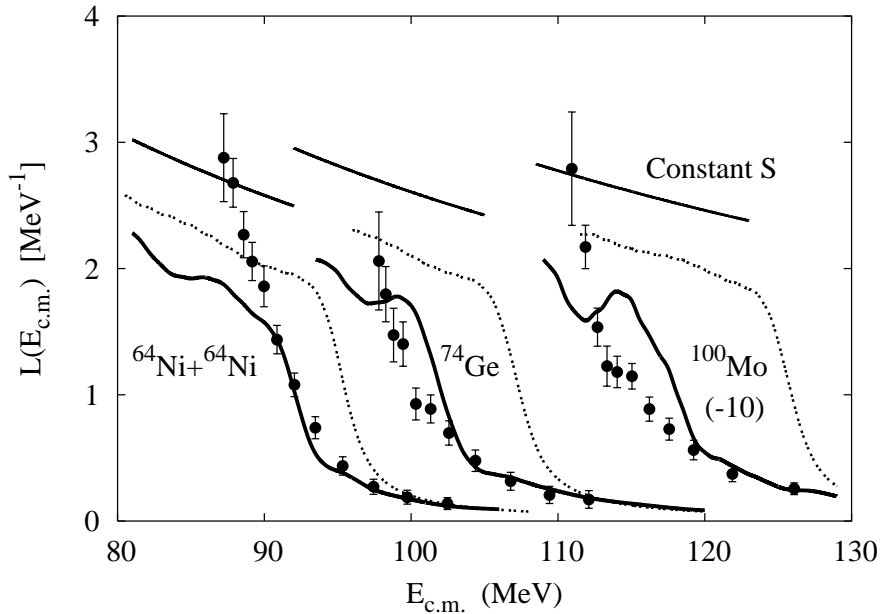}
\caption{\label{al1d} Logarithmic derivatives of the fusion cross sections 
shown in Fig. 1. The top lines show the constant {\it S} factor limit.} 
\end{figure}

\begin{figure}
\includegraphics[width =12cm]{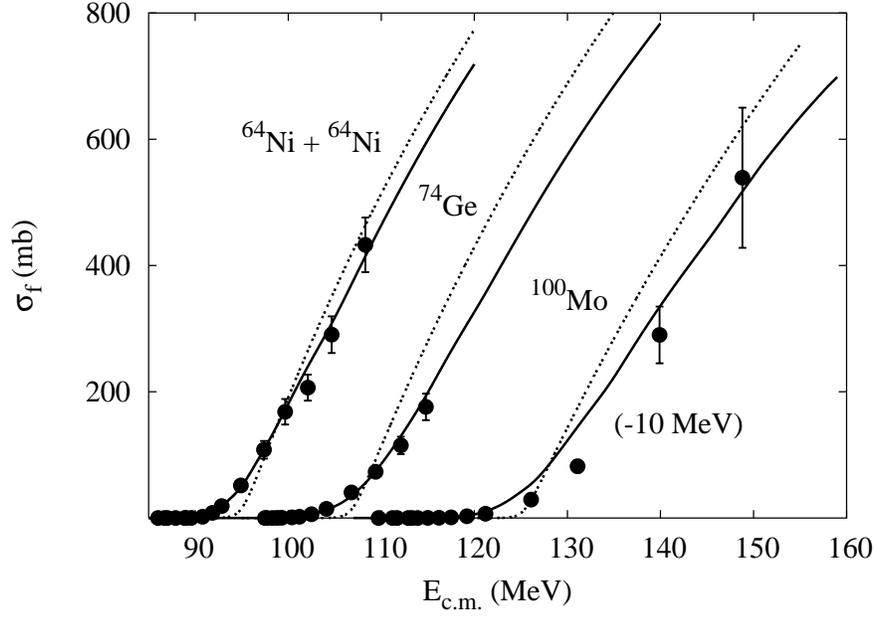}
\caption{\label{allf} Linear plot of the fusion cross sections shown in Fig. 1.}
\end{figure}

\begin{figure}
\includegraphics[width =12cm]{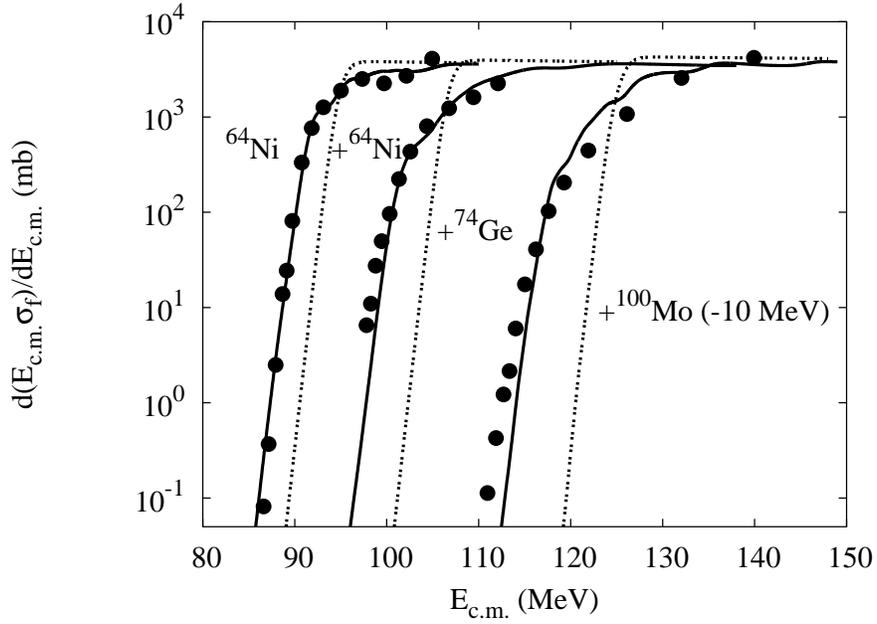}
\caption{\label{aldf} Derivative of $E_{c.m.}\sigma_f$, where the fusion cross 
sections $\sigma_f$ are shown in Fig. 1.}
\end{figure}

\begin{figure}
\includegraphics[width = 12cm]{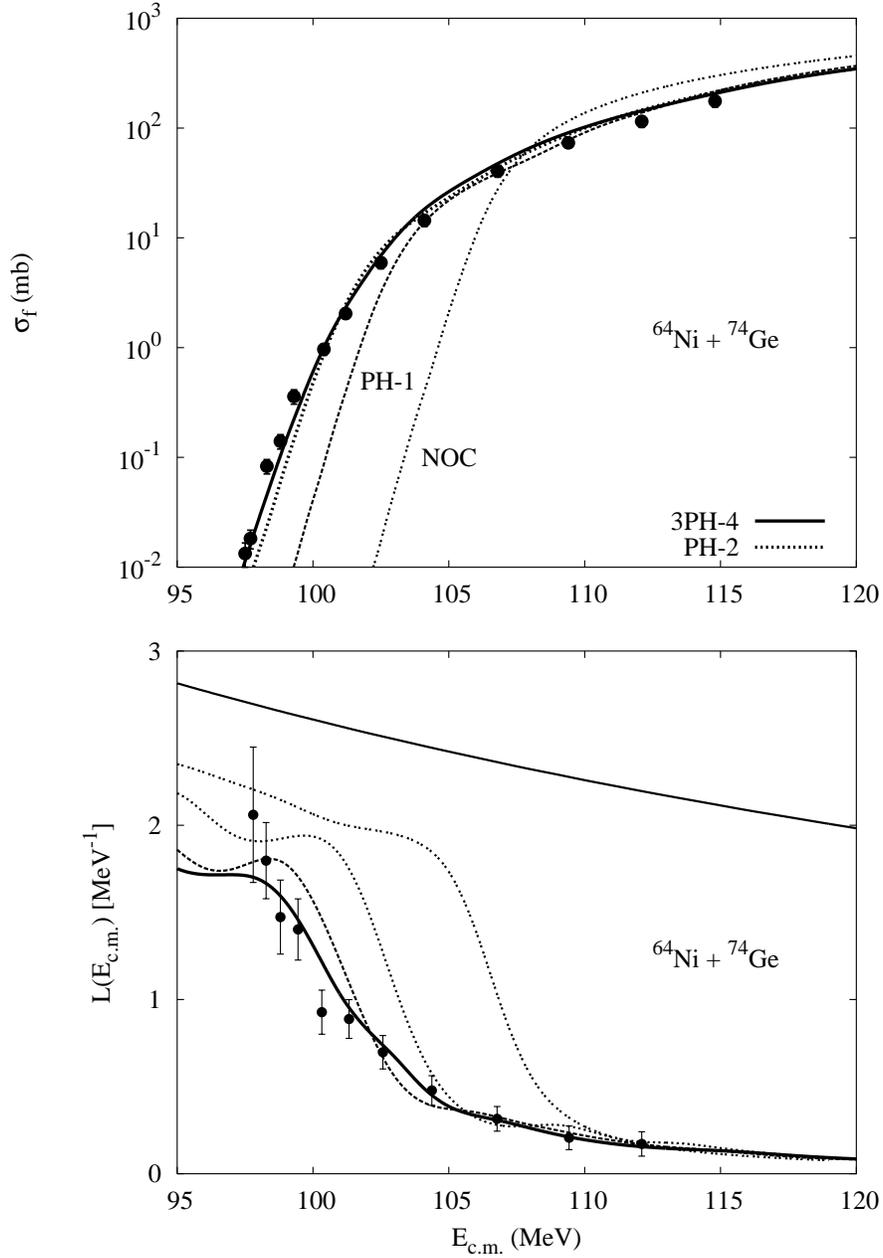}
\caption{\label{nigeco2} Fusion cross sections for $^{64}$Ni+$^{74}$Ge 
\cite{Beck} (top panel) and the associated logarithmic derivatives
(bottom panel). The dashed curves show the no-coupling limit (NOC),
the one-phonon (PH-1), and the basic two-phonon (PH-2) calculations.
The solid curves show the results of the 3PH-4 calculations.
All calculations used the radius parameter $\Delta R$=0.16 fm.
The upper line in the bottom panel is the result for constant $S$ factor.}
\end{figure}

\begin{figure}
\includegraphics[width = 12cm]{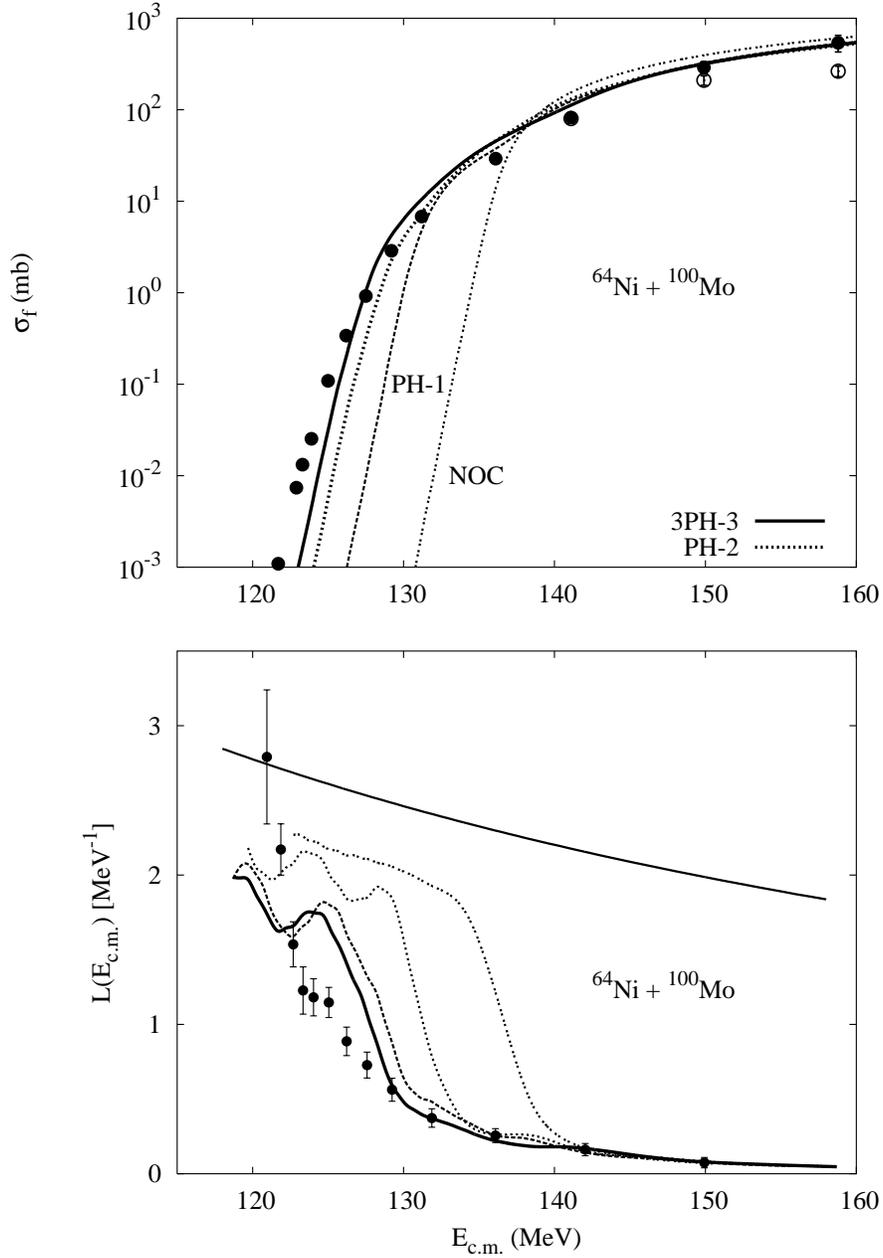}
\caption{\label{nimoco2} Fusion cross sections for $^{64}$Ni+$^{100}$Mo 
(top panel) and the associated logarithmic derivatives (bottom panel). 
The open circles are the measured evaporation cross section; the solid 
points include an estimated fission cross section \cite{nimo}.
The dashed curves show the no-coupling limit (NOC), the one-phonon 
(PH-1), and the basic two-phonon (PH-2) calculations.
The solid curves are the results of the 3PH-3 calculation.
All calculations used the radius parameter $\Delta R$ = 0.04 fm.
The upper line in the bottom panel is the result for constant $S$ factor.}
\end{figure}

\begin{figure}
\includegraphics[width = 12cm]{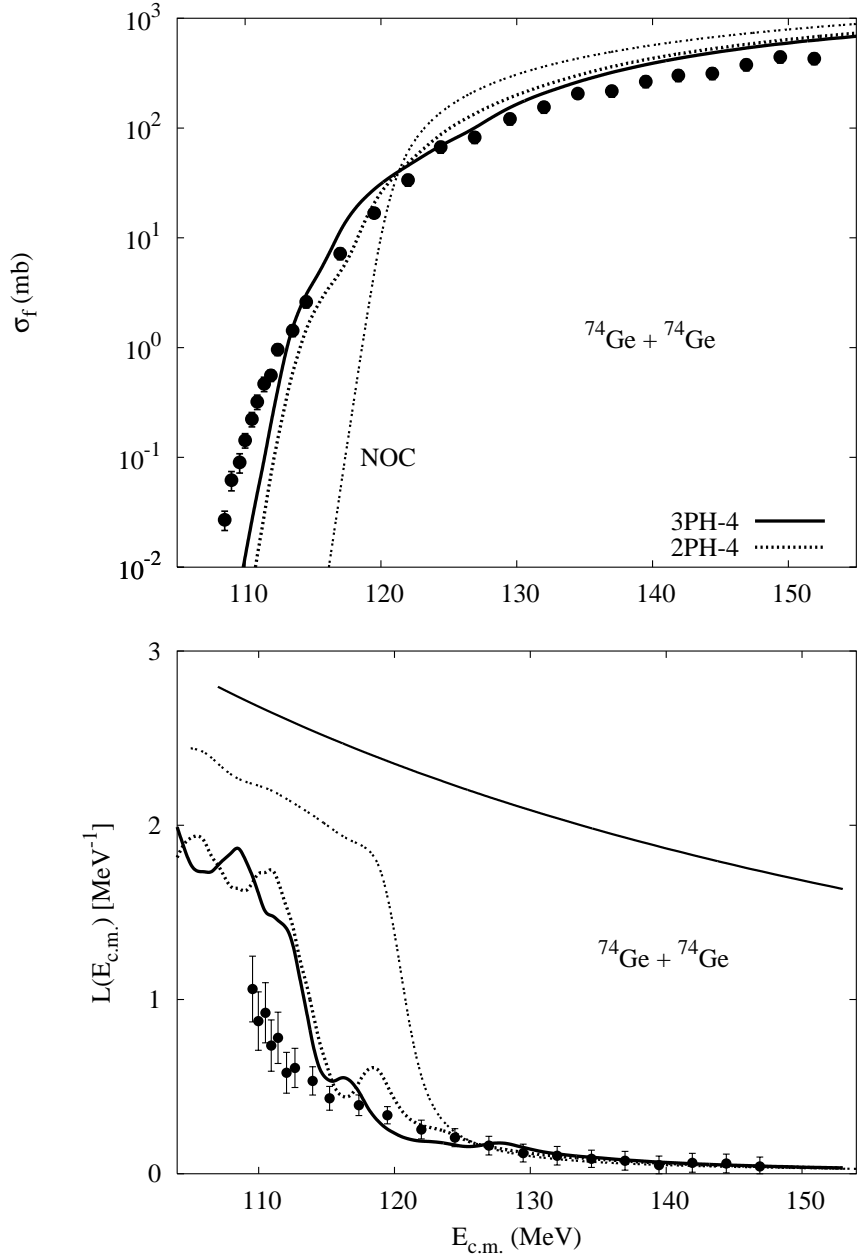}
\caption{\label{gegeco2} Fusion cross sections for $^{74}$Ge+$^{74}$Ge 
\cite{Beck} (top panel) and the associated logarithmic derivatives
(bottom panel). 
The upper line in the bottom panel is the result for constant $S$ factor.
The dashed curves show the no-coupling limit (NOC), and the 2PH-4
calculations. The solid curves show the result of the 3PH-4 calculation.
All calculations used the radius parameter $\Delta R$ = 0.10 fm. 
The upper line in the bottom panel is the result for constant $S$ factor.}
\end{figure}

\end{document}